\documentclass[sn-mathphys,Numbered]{sn-jnl}

\usepackage{amsmath,amssymb,amsfonts}
\usepackage{amsthm}
\usepackage{tikz}
\usepackage{braket}
\usepackage[ruled,vlined]{algorithm2e}
\usepackage{multirow}
\usepackage[normalem]{ulem}
\usepackage{graphicx}%
\usepackage{mathrsfs}%
\usepackage[title]{appendix}%
\usepackage{xcolor}%
\usepackage{manyfoot}%
\usepackage{booktabs}%
\newcommand\Vector[1]{\mathbf{#1}}

\newcommand\vk{{\Vector{k}}}
\newcommand\vq{{\Vector{q}}}
\newcommand\vr{{\Vector{r}}}

\newcommand\vR{{\Vector{R}}}

\newcommand\ik{{i\vk}}
\newcommand\iq{{i\vq}}

\newcommand\gO{\mathcal{O}}

\newcommand\upi{\mathord{\mathrm{i}}}

\newcommand\mpk{\mathcal{P}_k}

\theoremstyle{thmstyleone}%
\theoremstyle{thmstyletwo}%

\theoremstyle{thmstylethree}%

\begin{document}
\title[Article Title]{Hamiltonian transformation for accurate and efficient band structure interpolation}

\author[1,2]{\fnm{Kai} \sur{Wu}}

\author[3]{\fnm{Yingzhou} \sur{Li}}

\author[1]{\fnm{Wentiao} \sur{Wu}}

\author[4]{\fnm{Lin} \sur{Lin}}

\author*[1]{\fnm{Wei} \sur{Hu}}\email{whuustc@ustc.edu.cn}

\author*[1]{\fnm{Jinlong} \sur{Yang}}\email{jlyang@ustc.edu.cn}

\affil[1]{School of Future Technology, University of Science and Technology of China, Hefei, Anhui 230026, China}

\affil[2]{Dipartimento di Fisica, Università di Roma Tor Vergata, Via della Ricerca Scientifica 1, 00133 Rome, Italy}

\affil[3]{School of Mathematical Sciences, Fudan University, Shanghai 200433, China}

\affil[4]{Department of Mathematics, University of California, Berkeley, California 94720, United States}

\abstract{
Electronic band structure is a cornerstone of condensed matter physics and materials science.
Conventional methods like Wannier interpolation (WI), which are commonly used to interpolate band structures onto dense $\vk$-point grids, often encounter difficulties with complex systems, such as those involving entangled bands or topological obstructions. We introduce the Hamiltonian transformation (HT) method, a novel framework that enhances interpolation accuracy by localizing the Hamiltonian. Using a pre-optimized transformation, HT produces a far more localized Hamiltonian than WI-SCDM (where Wannier functions are generated via the selected columns of the density matrix projection), achieving up to two orders of magnitude greater accuracy for entangled bands. Although HT utilizes a slightly larger, nonlocal numerical basis set, its construction is rapid and requires no optimization, resulting in significant computational speedups. These features make HT a more precise, efficient, and robust alternative to WI-SCDM for band structure interpolation, as verified by high-throughput calculations.
}

\maketitle

\section{Introduction}\label{sec1}

The band structure is a fundamental concept in condensed matter physics and materials science, essential for predicting and understanding material properties and phenomena.
In the framework of Kohn-Sham density functional theory (DFT)~\cite{hohenberg1964inhomogeneous,kohn1965self}, band structure calculations typically involve three steps: (1) performing self-consistent field (SCF) electronic structure calculations on a uniform $\vk$-point grid $\{\vk\}$; (2) obtaining the Hamiltonian $H_\vq$ on a nonuniform $\vk$-point grid (or path) $\{\vq\}$; (3) diagonalizing $H_\vq$ to obtain eigenvalues.
Due to the complexity of the density functional, it is often more efficient to interpolate $H_\vq$ from $H_\vk$ in the second step using Fourier interpolation:
\begin{equation}
	H_\vq=\frac{1}{N_k}\sum_{\vk,\vR}H_\vk e^{\upi(\vq-\vk)\vR},
\end{equation}
where $\vR$ is the Bravais lattice vector, and $N_k$ is the number of uniform $\vk$-points.
In this paper, we focus on improving the accuracy of this interpolation.

The success of interpolation relies on the smoothness of matrix elements in reciprocal space or their localization in real space. To clarify, when we refer to the localization of the Hamiltonian, we mean localization in $\mathbf{R}$ space, not in the band indices $\alpha, \beta$. Specifically, for two unit cells located at $\mathbf{R}_i$ and $\mathbf{R}_j$, $H_{\alpha\beta}(\mathbf{R}_i, \mathbf{R}_j)$ decays to zero for sufficiently large $|\mathbf{R}_i - \mathbf{R}_j|$, regardless of the values of $\alpha$ and $\beta$. This is equivalent to $\|H(\mathbf{R}_i, \mathbf{R}_j)\|_2$ decaying to zero.  A faster decay means the Hamiltonian is more localized in real space.

Given that the DFT Hamiltonian is typically large, it must be projected onto a smaller basis set for practical interpolation. However, while the original implicit DFT Hamiltonian is localized in real space, the projected explicit smaller Hamiltonian is not necessarily so. This can result in a slow decay of the matrix elements with respect to $\vR$, necessitating a very large $N_k$ to achieve satisfactory interpolation accuracy. Thus, the challenge lies in constructing a small and localized Hamiltonian.

The maximally localized Wannier function (MLWF)\cite{marzari1997maximally, marzari2012maximally, pizzi2020wannier90} is a powerful tool widely used for interpolation, known as Wannier interpolation (WI). As a compact basis set, MLWFs are optimized to be as localized as possible, ensuring that the projected Hamiltonian remains localized. WI is a popular interpolation method in condensed matter physics and plays a crucial role in constructing model Hamiltonians \cite{jung2013tight,garrity2021database} and computing various physical observables of solids~\cite{wang2006ab,yates2007spectral,wang2017first}.
However, constructing MLWFs is a challenging nonlinear optimization problem due to the presence of multiple local minima\cite{marzari2012maximally}. Consequently, the results can be sensitive to initial guesses, requiring users to have detailed knowledge of the system to provide a good starting point.
Significant progress has been made in improving the robustness of numerical algorithms for finding localized Wannier functions~\cite{MustafaCohCohenEtAl2015,CancesLevittPanatiEtAl2017,StubbsWatsonLu2021,qiao2023projectability}.
One particularly robust approach is the selected columns of the density matrix (SCDM)~\cite{damle2015compressed,damle2017scdm,damle2018disentanglement}.
However, constructing MLWFs remains challenging in certain cases, such as topological insulators~\cite{soluyanov2011wannier,cornean2017wannier} and entangled band structures~\cite{SouzaMarzariVanderbilt2001,damle2018disentanglement,damle2019variational}.

Apparently, the Hamiltonian constructed from the ``maximally localized wavefunction" is not necessarily maximally localized. By instead optimizing with the localization of the Hamiltonian as the target function, we can obtain a truly ``maximally localized Hamiltonian".
In this work, we propose a new framework called Hamiltonian transformation (HT), specifically designed to directly localize the Hamiltonian.
Unlike MLWFs, HT does not involve any optimization procedure at runtime.
Instead, we design an invertible transform function $f$ that transforms Hamiltonian $H$ into $f(H)$, and optimize $f$ during the algorithm design phase to ensure $f(H)$ is as localized as possible. 
After diagonalizing $f(H)$ and obtaining the transformed eigenvalues $f(\varepsilon)$, the true eigenvalues can be recovered through the inverse transformation $\varepsilon=f^{-1}(f(\varepsilon))$.
Notably, the same transform function $f$ can also be applied within the WI framework, which yields an enhanced WI-SCDM-f scheme for more accurate model Hamiltonians.

HT offers two advantages over WI: (1) HT circumvents the complex optimization procedures required in WI by localizing the Hamiltonian through a pre-optimized transform function $f$, which we demonstrate to be universally applicable to all Hamiltonians; (2) By focusing on the localization of the Hamiltonian as the primary objective, HT achieves significantly higher accuracy (1 to 2 orders of magnitude better than WI-SCDM) in handling entangled bands.
We should note that HT has two disadvantages compared to WI: (1) HT cannot generate localized orbitals, which limits its ability to provide information about chemical bonds; (2) HT requires a larger basis set than WI, resulting in an interpolated Hamiltonian that is approximately an order of magnitude larger than that produced by WI.
In summary, the balance of advantages and limitations makes HT a specialized method for band structure interpolation: it is more accurate, more robust, and faster than WI-SCDM. HT is particularly effective for systems with entangled or topologically obstructed bands.

\section{Results}
\subsection{Designing the transform function $f$}

\begin{figure}[htbp]
	\begin{tikzpicture}
		\node[anchor=center] at (0,0) {\includegraphics[width=0.9\textwidth]{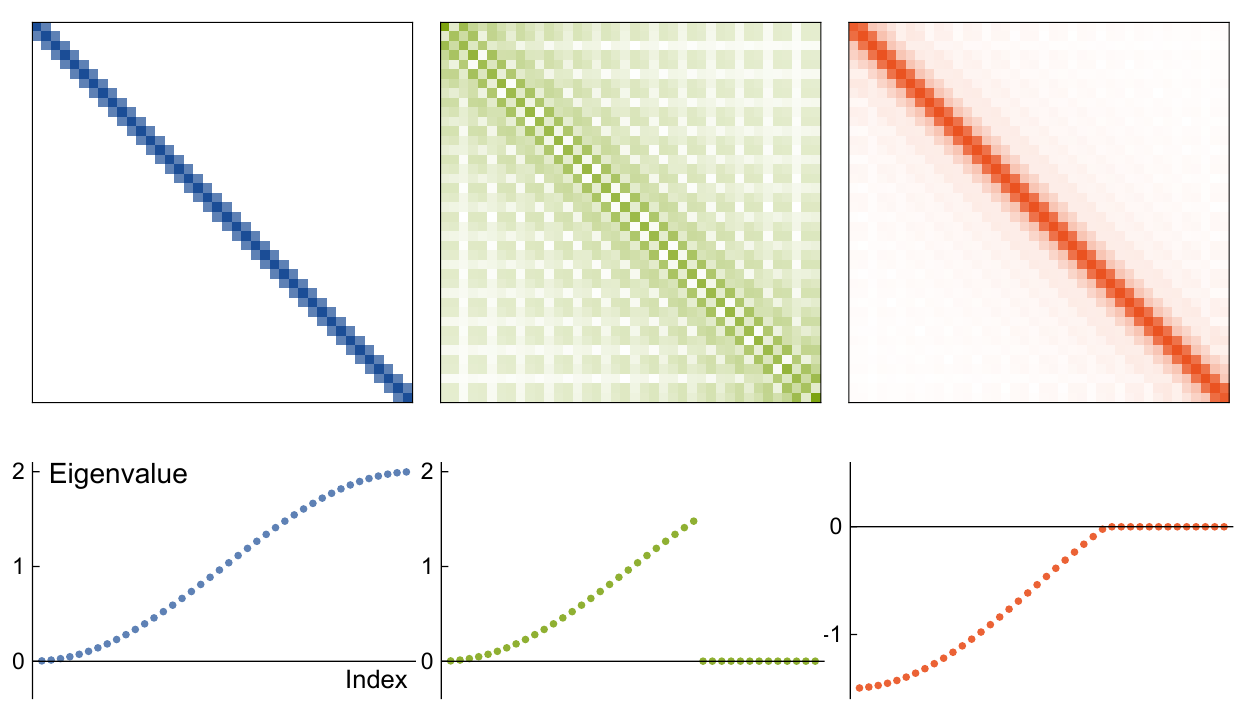}};
		\node[anchor=center] at (-5.6,3.3) {(a)};
		\node[anchor=center] at (-1.7,3.3) {(c)};
		\node[anchor=center] at (2.2,3.3) {(e)};
		\node[anchor=center] at (-5.6,-.8) {(b)};
		\node[anchor=center] at (-1.7,-.8) {(d)};
		\node[anchor=center] at (2.2,-.8) {(f)};
	\end{tikzpicture}
	\caption{An example demonstrating that modifying eigenvalues can recover the localization of the Hamiltonian. (a) Original tridiagonal Toeplitz Hamiltonian $T$ for a 1-D atomic chain with nearest-neighbor interactions.
		(b) Corresponding eigenvalue spectrum of $T$.
		(c) Reconstructed Hamiltonian after spectral truncation, leading to delocalization.
		(d) Truncated eigenvalue spectrum with eigenvalues below 1.5.
		(e) Reconstructed Hamiltonian after shifting the remaining eigenvalues downward by 1.5, showing improved localization.
		(f) Adjusted eigenvalue spectrum after the shift, restoring continuity.}
	\label{fig:eigenvalue}
\end{figure}

We begin with an example to demonstrate that the degradation of localization in the Hamiltonian is caused by spectral truncation. For a 1-D atomic chain with nearest-neighbor interactions, the Hamiltonian $T$ is a tridiagonal Toeplitz matrix~\cite{pasquini2006tridiagonal}. The main diagonal elements of $T$ are 1, and the lower and upper diagonal elements are 0.5, with all other elements being zero. The matrix $T$ and its eigenvalue spectrum are shown in Fig.~\ref{fig:eigenvalue}(a) and (b).
Although $T$ itself is localized, its eigenvectors are non-local, oscillating between positive and negative values, and canceling each other out away from the diagonal. In a typical SCF calculation, only a few of the lowest eigenvalues (assumed to be those less than 1.5 here) are obtained, corresponding to the truncated eigenvalue spectrum shown in Fig.\ref{fig:eigenvalue}(d). Reconstructing the Hamiltonian using only the truncated eigenvalues and eigenvectors results in a non-localized Hamiltonian, as shown in Fig.\ref{fig:eigenvalue}(c). After truncation, the eigenvalue spectrum becomes discontinuous, and the remaining eigenvectors are unable to cancel each other out effectively, leading to a delocalized reconstructed $T$.
A key observation is that by shifting the remaining eigenvalues downward by 1.5, we can restore continuity in the eigenvalue spectrum, as shown in Fig.\ref{fig:eigenvalue}(f). The reconstructed $T$ becomes significantly more localized, as illustrated in Fig.\ref{fig:eigenvalue}(e).

Therefore, the principle behind designing $f$ is to ensure that it smooths the eigenvalue spectrum. We will demonstrate later that optimizing $f$ is a multi-objective problem, making it difficult to determine the optimal form of $f$. A practical approach, therefore, is to design a family of $f$ functions with adjustable parameters and compare their effects.
The $f$ is designed by derivative:
\begin{equation}
	f_{a,n}'(x)=\begin{cases}
		0 & x \ge \varepsilon \\
		\frac{1}{2}-\frac{\text{erf}(n(\frac{1}{2}+\frac{x-\varepsilon}{a}))}{2\text{erf}(\frac{n}{2})} & \varepsilon-a\le x <\varepsilon\\
		1 & x < \varepsilon-a.
	\end{cases}\label{eq:f'}
\end{equation}
Here, $\varepsilon$ represents the maximum eigenvalue in the SCF calculation, and erf$(x)$ is the error function.
The function $f$ has two adjustable parameters, $a$ and $n$.
The parameter $a\ge 0$ controls the width of the transition region (typically set in proportion to the energy range of the entangled bands), while $n$ governs the smoothness of the function $f$; a larger $n$ results in a smoother function.
The formula of $f$ is obtained by integral from $f'$ with $f(\varepsilon)=0$, which is shown in Eq.~\eqref{eq:f}.
\begin{equation}
	f_{a,n}(x)=\begin{cases}
		0 & x\ge\varepsilon\\
		\frac{\frac{2 a (e^{-\frac{n^2}{4}}-e^{-\frac{n^2 (2 x+a)^2}{4 a^2}})}{\sqrt{\pi } n}+(2 x+a) \left(\text{erf}\left(\frac{n}{2}\right)-\text{erf}\left(n\left(\frac{x}{a}+\frac{1}{2}\right)\right)\right)}{4 \text{erf}\left(\frac{n}{2}\right)} & \varepsilon-a\leq x< \varepsilon \\
		x+a/2 & x<\varepsilon-a
	\end{cases}.\label{eq:f}
\end{equation}

\begin{figure}[htbp]
	\begin{tikzpicture}
		\node[anchor=center] at (0,0) {\includegraphics[width=0.95\textwidth]{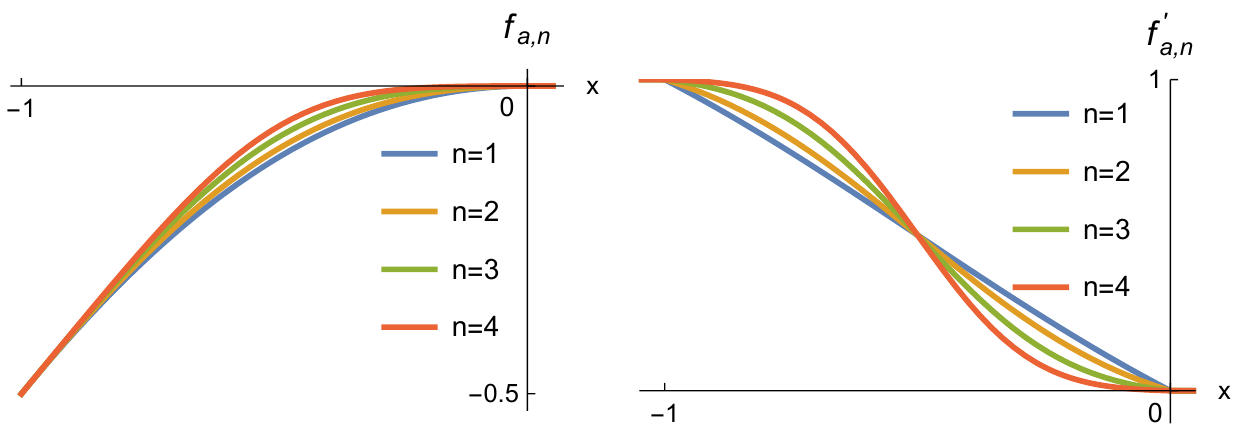}};
		\node[anchor=center] at (-6.1,2) {(a)};
		\node[anchor=center] at (0.2,2) {(b)};
	\end{tikzpicture}
	\caption{The transform function and its derivative. (a) The transform function $f_{a,n}(x)$ for different values of $n$, with the transition region width $a=1$. As $n$ increases, $f_{a,n}(x)$ becomes smoother. (b) The derivative $f_{a,n}'(x)$. Higher values of $n$ result in a more gradual change in slope.}
	\label{fig:transform}
\end{figure}

Without loss of generality, we assume $\varepsilon=0$ in the following discussion. The plots of $f_{a=1,n}(x)$ and $f_{a=1,n}'(x)$ are shown in Fig.~\ref{fig:transform}(a) and (b), respectively.
In Fig.~\ref{fig:transform}(a), the piecewise function $f_{a,n}(x)$ consists of three parts: the right part, for $x>0$, where $f_{a,n}(x)$ is set to 0, simulating the truncation of eigenvalues; the left part, for $x<-1$, which is linear, ensuring that eigenvalues significantly less than 0 undergo only a constant shift; and the middle part, which acts as a smoother, providing a gradual transition between the two linear regions.

\subsection{Localization functional $F$}
In this section, we introduce a functional $F$ to quantitatively describe the localization properties of any sparse Hermitian Hamiltonian.
In the plane-wave basis set, the DFT Hamiltonian is generally assumed to be a dense matrix. However, to achieve more accurate interpolation, we must adopt a sufficiently large $\mathbf{k}$-point mesh, which is equivalent to using a larger supercell in real space. This enlargement ensures that for the farthest two unit cells, $\vR_i$ and $\vR_j$, $\|H(\mathbf{R}_i, \mathbf{R}_j)\|_2$ becomes sufficiently small, avoiding overlap with periodic mirror images. In this case, the Hamiltonian effectively becomes a sparse matrix.

The basic approach to analyzing the decay properties of a sparse matrix involves approximating the transform function using polynomials and analyzing the expansion coefficients. Similar ideas have been applied to study the sparsity of density matrices~\cite{baer1997sparsity, benzi2013decay}.

In the following discussion, we assume the band indices $\alpha$, $\beta$ of Hamiltonian are fixed, thereby omitting them and simplifying $H_{\alpha\beta}(\mathbf{R}_i, \mathbf{R}_j)$ to $H_{ij}$.
Consider an $m$-banded Hermitian matrix $H$ with the following properties: (1) The eigenvalue spectrum $\sigma(H)$ lies within the interval $[-1,1]$ (if not, $H$ can be scaled to meet this requirement); (2) There exists an integer $m \ge 0$ such that $H_{ij} = 0$ when $|i - j| > m$.
We define the $k$-th best approximation error of a continuous transform
function $f$ on the closed interval $[-1,1]$ (i.e. $f\in C[-1,1]$) as
\begin{equation} \label{eq:E_k}
	E_k(f) = \inf\left\{ \max_{-1\le x\le 1}|f(x)-p(x)|: p\in\mpk \right\},
\end{equation}
where $\mpk$ denotes the subspace of algebraic polynomials of degree at most
$k$ in $C[-1,1]$.
Let $i,j$ indices satisfy $mk<|i-j|\le m(k+1)$, for any $p_k \in \mathcal{P}_k$, we have $p_k(H)_{ij}=0$.
Thus
\begin{equation}
	\begin{split}
		\left|f(H)_{ij}\right| & = \left|[f(H)-p_k(H)]_{ij}\right|\\
		& \le \left\|f(H)-p_k(H)\right\|_2 = \max_{x\in\sigma(H)}\left|f(x)-p_k(x)\right|\\
		& \le \max_{-1\le x\le 1}\left|f(x)-p_k(x)\right|,
	\end{split}\label{eq:inequality}
\end{equation}
which means that
\begin{equation}
	\left|f(H)_{ij}\right| \le E_k(f).
\end{equation}
In Eq.~\eqref{eq:inequality} we have used
\begin{equation}
	|A_{ij}|\le\sqrt{\sum_i|A_{ij}|^2}=\left\|Ae_j\right\|_2\le\sup_{x\ne\mathbf{0}}\frac{\left\|Ax\right\|_2}{\left\|x\right\|_2}=\left\|A\right\|_2.
\end{equation}

The exact expression for the optimal $p_k$ is unknown, but we can
approximate $E_k(f)$ using Chebyshev polynomials. Approximation
theory guarantees that Chebyshev polynomials are nearly optimal, and error
bounds for the Chebyshev series are well-established for smooth
functions~\cite{bernstein1912ordre, xiang2010error}.
Here we calculate exact error bounds for certain specific functions.

The expression of $f$ in terms of the Chebyshev polynomial basis is given by:
\begin{equation}
	f(x)=\frac{1}{2}\alpha_0+\sum_{l=1}^{\infty}\alpha_l T_l(x), \label{eq:decompose}
\end{equation}
\begin{equation}
	\alpha_l=\frac{2}{\pi}\int_{0}^{\pi}f(\cos \theta)\cos{l\theta} d\theta,\label{eq:alpha}
\end{equation}
where $T_l(x)$ is the $l$th Chebyshev polynomial of the first kind.
As a result, the decay properties of $f(H)$ can be estimated by
\begin{equation}
	\begin{split}
		&|f(H)_{ij}|\le E_k(f)\le\left\|\sum_{l=k+1}^{\infty}\alpha_{l}T_l(x)\right\|_{x\in[-1,1]}\\
		=&\frac{2}{\pi}\left\|\sum_{l=k+1}^{\infty}\cos{l\theta}\int_{0}^{\pi}f(\cos t)\cos{l t}\ dt\right\|_{\theta\in[0,\pi]}\\
		=&c\ F[f,k],
	\end{split}\label{eq:F}
\end{equation}
where $c$ is a factor normalizing $F[f,0]$ to 1.

Up to this point, we have obtained a functional $F$ in Eq.~\eqref{eq:F} to analyze the localization properties of Hamiltonian. An explanation of $F$ is that, for any banded Hermitian matrix $H$ with bandwidth $m$ and eigenvalues in $[-1, 1]$, if we apply a transformation $f$ to $H$, then $|f(H)_{ij}|$ is bounded above by $cF[f,k]$, where $k$ is an integer satisfying $mk < |i - j| \leq m(k + 1)$.
Although $H$ is restricted to a banded matrix, the results presented in this section can be extended to general sparse matrices, provided that $H$ is associated with a sparsely connected, degree-limited graph~\cite{benzi2013decay}.

\subsection{Optimizing transform function $f_{a,n}$}
\begin{figure}[htbp]
	\begin{tikzpicture}
		\node[anchor=center] at (0,0) {\includegraphics[width=\textwidth]{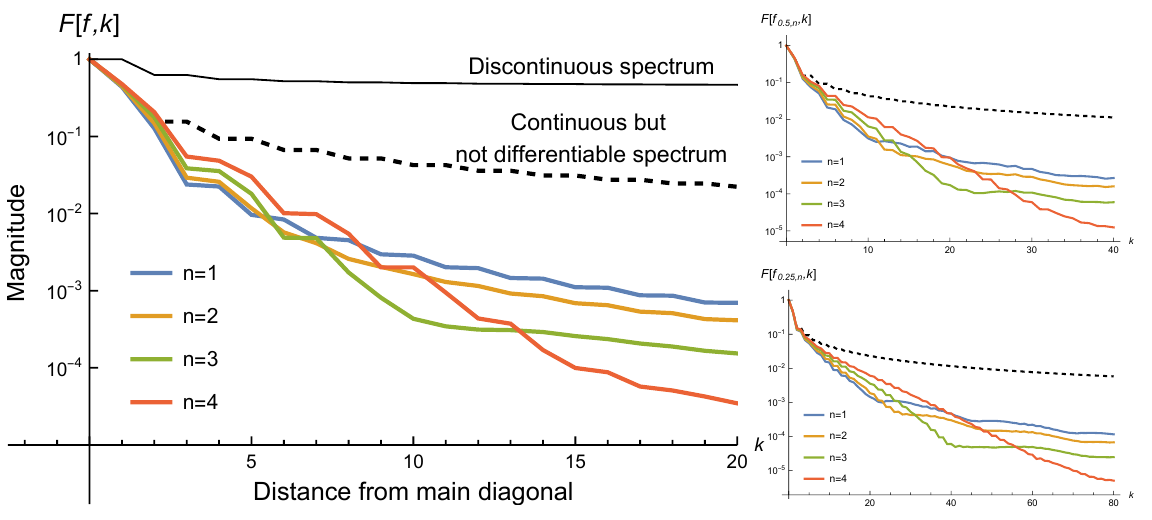}};
		\node[anchor=center] at (-6.3,2.8) {(a)};
		\node[anchor=center] at (1.86,2.75) {(b)};
		\node[anchor=center] at (1.86,-.1) {(c)};
	\end{tikzpicture}
	\caption{Decay of off-diagonal elements of transformed Hamiltonian.
		(a) Decay properties of the $m$-banded Hermitian matrix $H$ after transformation, $|f_{a,n}(H)_{ij}|\le
		c_{a,n}F[f_{a,n},k]$, $mk<|i-j|\le m(k+1)$, $c_{a,n}$ is a factor
		normalizes $F[f_{a,n},0]$ to 1. We emphasize that the results apply to \textbf{all $m$-banded Hermitian matrices}. (b) and (c) show similar decay behavior as in (a), but with the transition region width $a$ set to 0.5 and 0.25, respectively.}
	\label{fig:F[f]}
\end{figure}

By substituting $f_{a,n}$ from Eq.~\eqref{eq:f} into $F$ in Eq.~\eqref{eq:F}, and using Eq.~\eqref{eq:decompose} to simplify $\sum_{k+1}^\infty$ to $\sum_{1}^k$, we obtain the numerical results shown in Fig.~\ref{fig:F[f]}.

In Fig.~\ref{fig:F[f]}(a), the black solid line corresponds to the case where $f(x) = \Theta(-x)(x - 0.5)$, simulating a discontinuous eigenvalue spectrum with a gap of 0.5. This line does not decay to zero, indicating that, in some extreme cases, for the farthest two unit cells located at $\mathbf{R}_i$ and $\mathbf{R}_j$, $\|H(\mathbf{R}_i, \mathbf{R}_j)\|_2$ converges to a nonzero value as $N_k \to \infty$.
The black dashed line represents $F[f_{0,n},k]$, which corresponds to a continuous but non-differentiable spectrum. It decays rapidly for $k\le2$, but more slowly for larger $k$. The colored solid lines in  Fig.~\ref{fig:F[f]}(a) represent $F[f_{1,n},k]$. These lines decay significantly faster than the black dashed line, indicating that the transform function $f_{1,n}$ is more effective than merely shifting the eigenvalues.
Fig.~\ref{fig:F[f]}(b) and Fig.~\ref{fig:F[f]}(c) show plots where the transition region width $a$ is set to 0.5 and 0.25, respectively. These figures display similar behavior to the $a=1$ case after rescaling, with larger $a$ leading to faster decay of $F$.

There are two considerations when choosing the parameters $a$ and $n$. First, each colored line in Fig.~\ref{fig:F[f]} exhibits an inflection point where $F$ transitions from rapid to slower decrease. With small $n$, $F$ decays quickly initially but reaches the inflection point early, leading to slower decay afterward. Conversely, larger $n$ values result in a slightly slower initial decay but delay the inflection point, causing $F$ to decay faster when $k$ is sufficiently large. Second, for large $a$ and $n$, the inverse function $f_{a,n}^{-1}(x)$ becomes ill-conditioned near $x = 0$, introducing more errors in the top bands. This necessitates including more bands in the SCF calculations. Based on our experience, setting $n = 3$ and
\begin{equation}
	a=4(\max_{\vk}(\varepsilon_\ik)-\min_{\vk}(\varepsilon_\ik)),
\end{equation}
where $i$ is the index of the top band, provides a good balance between decay rate and the number of bands required for interpolation.
Unless otherwise specified, our simulations will use this set of parameters.
Further details regarding the choice of the parameter \(n\) are provided in the Results section.

\subsection{Basis set transformation}
The DFT Hamiltonian is usually too large to interpolate directly. We reduce the size of the Hamiltonian by changing to a relatively small, $\vk$-independent numerical basis set:
\begin{equation}
	\psi_\ik(\vr)=\sum_{\mu=1}^{N_\mu} Q_\mu(\vr)C_{\mu,\ik}.\label{eq:psiexpand}
\end{equation}
Here, $N_\mu$ is the size of basis set, $\psi_\ik(\vr)=e^{\upi \vk\cdot \vr}u_\ik(\vr)$ is the Bloch wavefunction in real space, and $u_{i\vk}(\vr)$ is the periodic part within the unit cell.
In Eq.~\eqref{eq:psiexpand}, decomposition is performed within the unit cell at $\vR = \mathbf{0}$, not the entire supercell.
By using the basis set $Q$, we can perform Fourier interpolation on a smaller $N_\mu \times N_\mu$ matrix, making the process more efficient.

The simplest method to perform such decomposition is singular value decomposition (SVD), but it is slow in large basis set.
A specialized algorithm for this task is developed based on randomized QR factorization with column pivoting (QRCP)\cite{duersch2017randomized}, with technical details provided in Supplemental Material S1.
Randomized QRCP is highly efficient, accounting for only a small fraction of the total computational time.

Compared to MLWFs, the basis functions $Q_\mu(\vr)$ are independent of $\vk$, meaning that orbitals at all $\vk$-points share the same auxiliary basis. Changing to this basis set does not affect the decay properties of the Hamiltonian. On the other hand, a disadvantage of using $Q_\mu(\vr)$ is that they are non-localized and cannot provide information about chemical bonds.
Additionally, the size of this basis set is typically one order of magnitude larger than that of the Wannier basis set.

\subsection{Hamiltonian transformation and time complexity}\label{sec:complexity}
By combining the eigenvalue transformation function $f$ with the change of basis set, we propose the Hamiltonian Transformation (HT) method to interpolate physical quantities such as the band structure. This method is outlined in Algorithm~\ref{al:ht}.

We constructs the numerical basis $Q_\mu(\vr)$ from DFT orbitals $\psi_{i\vk}(\vr)$ obtained on a uniform $\vk$-grid.
To handle nonorthogonal orbitals from the projector augmented wave (PAW) method or ultrasoft pseudopotentials, HT computes the overlap matrix $\tilde{S}_{\mu\nu}$ in the basis $Q_\mu(\vr)$ and builds the Hamiltonian $H_{\vk}$ using the coefficients $\tilde{C}_{\mu,i\vk}$.
An eigenvalue transform $f$ then produces $f(H_\vk)$ with enhanced real-space locality, which is Fourier-interpolated to the desired $\vq$-points. Finally, HT solves the generalized eigenproblem for $f(H_\vq)$ with $\tilde{S}_{\mu\nu}$ and recovers the true eigenvalues via the inverse transform $f^{-1}$.

\begin{algorithm}
	\caption{Hamiltonian transformation for band structure calculation}\label{al:ht}
	\SetKwInOut{KwInput}{Input}
	\SetKwInOut{KwOutput}{Output}
	\KwInput{uniform grid $\{\vk\}$, nonuniform path $\{\vq\}$,\\
		eigenvalues $\{\varepsilon_\ik\}$, eigenvectors $\{\psi_\ik(\vr)\}$, overlap matrix $S(\vr,\vr')$}
	\KwOutput{$\{\varepsilon_\iq\}$}
	1. Construct the numerical basis set\;
	\quad $\psi_{i\vk}(\vr)=\sum_\mu Q_\mu(\vr)C_{\mu,i\vk}$\;
	2. Construct the explicit Hamiltonian\;
	\quad $\tilde{S}_{\mu\nu}=\int d\vr d\vr' Q^*_{\mu}(\vr)S(\vr,\vr')Q_{\nu}(\vr')$\;
	\quad $\tilde{C}_{\nu,i\vk}=\sum_{\mu}\tilde{S}_{\nu\mu}C_{\mu,i\vk}$\;
	\quad $f(H_{\vk,\mu\nu})=\sum_{i}f(\varepsilon_\ik) \tilde{C}_{\mu,\ik} \tilde{C}_{\nu,\ik}^*$\;
	3. Fourier interpolate the Hamiltonian\;
	\quad $f(H_{\vq,\mu\nu})=\frac{1}{N_k}\sum_{\vk,\vR}f(H_{\vk,\mu\nu}) e^{\upi(\vk-\vq)\cdot\vR}$\;
	4. Diagonalize the interpolated Hamiltonian\;
	\quad $f(H_{\vq,\mu\nu})=\sum_{i}f(\varepsilon_\iq) \tilde{C}_{\mu,\iq} \tilde{C}_{\nu,\iq}^*$\;
	5. Recover the eigenvalues\;
	\quad $\varepsilon_\iq=f^{-1}(f(\varepsilon_\iq))$\;
\end{algorithm}

\subsection{High-throughput accuracy tests}
To verify the effectiveness of HT and compare it with WI, we perform high-throughput calculations using a database~\cite{vitale2020automated} containing 200 materials that span a wide range of structural and chemical spaces.
Among these materials, 187 have at least 6 bands around the Fermi level with entangled band structures and are selected for our tests.
We use the SCDM method to construct MLWFs within the WI framework.
The free parameters in the SCDM method are determined using an automatic projection procedure~\cite{vitale2020automated,qiao2023projectability}.
To evaluate the interpolation accuracy, we exclude the highest $m$ bands and calculate the mean absolute error (MAE) of the remaining eigenvalues using:
\begin{equation}
	\text{MAE}=\frac{\sum_{i=1}^{N_b-m}\sum_\vk|\varepsilon_\ik^\text{interpolation}-\varepsilon_\ik^\text{benchmark}|}{N_k(N_b-m)}.
\end{equation}
In our calculations, we set $m=4$ and use the non-self-consistent field (non-SCF) DFT band structures as the benchmark. Besides HT and WI-SCDM, we also test a combined approach where we apply the transformation function within the WI-SCDM method. Specifically, we transform the eigenvalues before applying WI-SCDM and then transform them back after the interpolation.
We set $n=3$ for the transform function $f$ and refer to this method as ``WI-SCDM-f".

\begin{figure*}[htbp]
	\begin{tikzpicture}
		\node[anchor=center] at (0,0) {\includegraphics[width=\textwidth]{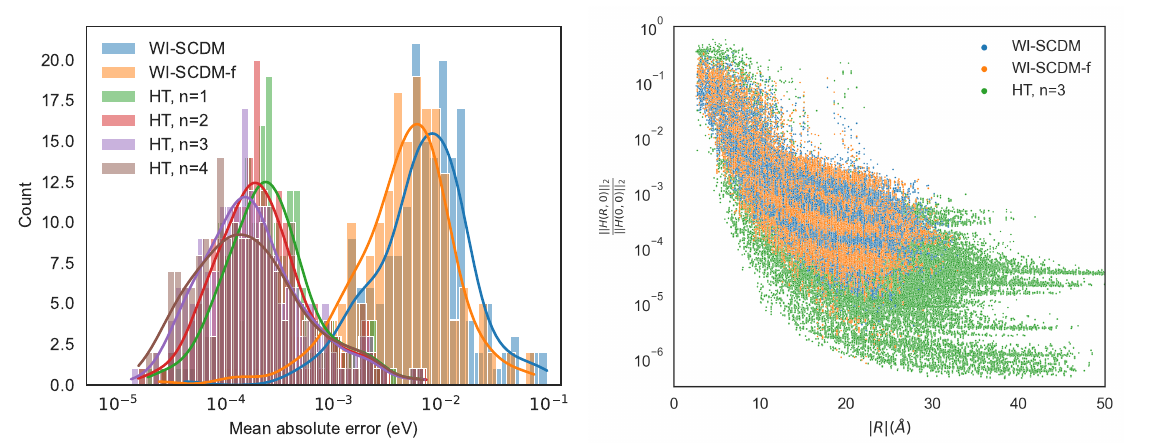}};
		\node[anchor=center] at (-6.3,2.5) {(a)};
		\node[anchor=center] at (.3,2.5) {(b)};
	\end{tikzpicture}
	\caption{High-throughput mean absolute error (MAE) distribution and Hamiltonian decay behavior.
		(a)Histogram of MAEs for WI-SCDM, WI-SCDM-f, and HT with $n=1-4$ across 187 materials with entangled bands.
		HT yields the lowest MAEs, with the distribution shifting left as $n$ increases, and outperforms WI-SCDM and WI-SCDM-f methods.
		The largest HT error is $7\times10^{-3}$~eV for CBe$_2$, which is further analyzed in Fig.~\ref{fig:accuracy}(a).
		(b) Decay properties of Hamiltonians in high-throughput calculations. Generally, HT Hamiltonians exhibit faster decay than WI-SCDM and WI-SCDM-f Hamiltonians.}\label{fig:high}
\end{figure*}

We compute the entangled band structures from the database using WI-SCDM, WI-SCDM-f, and HT, then calculate the MAE of the interpolated eigenvalues and present the cumulative frequency histogram of the MAE in Fig.~\ref{fig:high}(a).
The x-axis displays the MAE on a logarithmic scale from $10^{-5}$ to $10^{-1}$, and the y-axis shows the frequency (count) of occurrences for each error magnitude.
The overall distribution for each method forms a peak, emphasized by an envelope curve.
WI-SCDM exhibits the largest errors, with its peak around $10^{-2}$ eV.
Through eigenvalue transformation, WI-SCDM-f slightly outperforms WI-SCDM, demonstrating that incorporating $f$ into the WI-SCDM workflow yields more accurate model Hamiltonians.
HT, however, significantly outperforms both, with its peak around $10^{-4}$ eV, indicating much lower errors.
We also study the effect of \(n\) of the transform function \(f\). As \(n\) increases from 1 to 4, the peak of the HT error distribution shifts progressively leftward. The largest improvement occurs between \(n=1\) and \(n=3\), with diminishing returns beyond \(n=3\), suggesting a practical optimum at \(n=3\).

Furthermore, we present the decay properties of the Hamiltonians from high-throughput calculations in Fig.~\ref{fig:high}(b).
The x-axis represents $|\vR|$, and the y-axis shows $||H(\vR,0)||_2/||H(0,0)||_2$, indicating the relative strength of Hamiltonian elements as a function of distance.
Since we are interpolating entangled band structures, the Hamiltonian elements do not decay exponentially but rather exhibit an initial rapid decay within the first 20-30 \AA, followed by a slower, long-range decay.
The WI-SCDM and WI-SCDM-f tight-binding Hamiltonians are projected onto coarser $\vk$-point grids, resulting in fewer data points compared to the HT Hamiltonians.
Both WI-SCDM and WI-SCDM-f Hamiltonians display a similar decay trend, with values ranging from $10^{-5}$ to $10^{-3}$ when $|\vR| = 20$ \AA. In contrast, the HT Hamiltonians show a wider spread, ranging from $10^{-6}$ to $10^{-3}$ at $|\vR| = 20$ \AA. Overall, we observe that the HT Hamiltonians exhibit the fastest decay rate.

\begin{figure*}[htbp]
	\centering
	\begin{tikzpicture}
		\node[anchor=center] at (0,0) {\includegraphics[width=\textwidth]{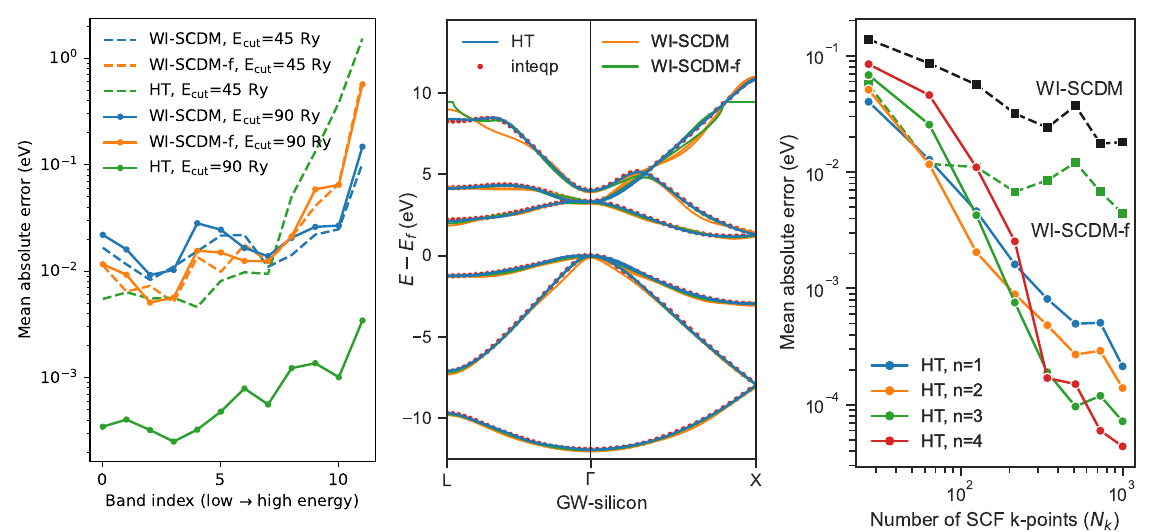}};
		\node[anchor=center] at (-6.2,2.9) {(a)};
		\node[anchor=center] at (-1.8,2.9) {(b)};
		\node[anchor=center] at (2.7,2.9) {(c)};
	\end{tikzpicture}
	\caption{Case studies of interpolation accuracy and k-point convergence.
		(a) Band‐resolved MAE for CBe$_2$, the material for which HT exhibits the largest errors in our dataset. Increasing the plane‐wave cutoff energy $E_\text{cut}$ from 45~Ry to 90~Ry dramatically reduces the HT MAE, while the MAEs of WI‐SCDM and WI‐SCDM‐f remain essentially unchanged.
		(b) GW quasiparticle band structures for silicon, with HT showing the best agreement with the benchmark of inteqp. An extremely sparse k-point mesh is used here, and the significant errors in WI-SCDM and WI-SCDM-f indicate they require a much larger $N_k$ to achieve sufficient accuracy.
		(c) MAE of silicon as a function of $N_k$. HT outperforms WI-SCDM and WI-SCDM-f, with its error rapidly decreasing as $N_k$ increases.
	}\label{fig:accuracy}
\end{figure*}

To further analyze the performance of HT, we focus on CBe$_2$, where HT exhibits the largest MAE among all 187 structures.
Figure~\ref{fig:accuracy}(a) shows the band-resolved MAE distribution for CBe$_2$. In the high-throughput calculation of Fig.~\ref{fig:high}(a), using a plane-wave cutoff energy $E_\text{cut}$ of 45~Ry, HT reaches an MAE of $7\times10^{-3}$~eV.
Increasing $E_\mathrm{cut}$ to 90~Ry reduces the MAE of HT to below $10^{-3}$~eV for most bands.
In contrast, both WI-SCDM and WI-SCDM-f show negligible change with $E_\text{cut}$, indicating their dominant error arises from the disentanglement procedure rather than plane-wave convergence.
Therefore, the poor performance of HT on some materials is primarily due to insufficient cutoff energy.
Raising $E_\text{cut}$ significantly reduces the interpolation error.

We also observe that the MAE in Fig.~\ref{fig:accuracy}(a) increases with the band index, where the highest-energy bands showing the largest errors. Such a band-dependent behavior arises from two factors: (1) the top bands are entangled with higher-energy bands that are excluded from the interpolation; (2) in HT, the slope of $f$ vanishes near these bands, making the inverse transform $f^{-1}$ ill-conditioned in that region. This issue can be mitigated by including additional bands in the calculation and discarding them after interpolation.

Unlike the DFT Hamiltonian, the GW quasiparticle Hamiltonian is more non-local.
We perform calculations on Si$_2$ to compare the performance of different methods.
To make the interpolation errors more apparent, we intentionally chose a very sparse $\vk$-point mesh ($5 \times 5 \times 5$). The results are shown in Fig.~\ref{fig:accuracy}(b).
The red points represent benchmarks obtained using the inteqp method from BerkeleyGW\cite{deslippe2012berkeleygw}, which requires additional information (the orbitals on fine $\vk$-point grids) compared to WI-SCDM and HT.
The WI-SCDM results (orange lines) display visible errors, but these errors are reduced after applying the transformation (green lines). The HT band structures (blue lines) show the best agreement with the red benchmark points.
It should be noted that the errors shown in Fig.~\ref{fig:accuracy}(b) do not indicate failure of the two WI-based methods; rather, they merely require a significantly larger $N_k$ to achieve comparable accuracy.

We test the accuracy of HT and WI-SCDM with respect to $N_k$ by performing DFT calculations on silicon, increasing $N_k$, and comparing their MAEs for the lowest 8 bands along the path between $\Gamma$ and X.
The results are shown in Fig.\ref{fig:accuracy}(c).
We observe that WI-SCDM exhibits the lowest accuracy, and introducing the transformation function improves its performance. However, both methods encounter a bottleneck: when $N_k$ reaches a certain threshold, their MAEs decrease much more slowly and begin to oscillate.
In contrast, HT is more accurate than both WI-SCDM and WI-SCDM-f, and its accuracy can be systematically improved by increasing $N_k$.
Furthermore, the MAEs of HT in Fig.\ref{fig:accuracy}(c) display decay patterns similar to those of the lines in Fig.~\ref{fig:F[f]}(a).
Specifically, when $N_k$ is small, a smaller $n$ leads to a smaller MAE, whereas when $N_k$ is large, a larger $n$ results in a smaller MAE. This similarity further verifies the theoretical results.

\subsection{Computational time scaling and performance}
The theoretical time complexity of HT is shown in Table~\ref{tab:complexity}.
Here, $N_r$ represents the number of real space grids, $N_\mu$ is the size of the new basis set, and $N_k$ is the number of SCF $\mathbf{k}$-points. Additionally, $N_b$ and $N_q$ denote the number of bands and the number of $\mathbf{k}$-points in the band structure calculation, respectively.
Assuming that $N_r$, $N_\mu$, and $N_b$ are proportional to the number of electrons $N_e$, and $N_q$ remains constant, the total time complexity of HT is $\mathcal{O}\left( N_e^3 N_k \log(N_k) \right)$.
HT and WI share the same time complexity, but their speed differs due to two factors: HT does not rely on run-time optimization, while WI uses a smaller basis set.

\begin{table}[htbp]
	\caption{ Theoretical time complexity of various procedures in Hamiltonian transformation. $N_r$: number of real space grids, $N_\mu$: size of new basis set, $N_k$: number of SCF $\mathbf{k}$-points, $N_b$: number of bands, and $N_q$: number of $\mathbf{k}$-points in the band structure calculation.}
	\begin{tabular*}{\linewidth}{c@{\extracolsep{\fill}}cc} \\ \hline \hline
		Operation & Algorithm & Time complexity \\ \hline
		Construct basis set & Randomized QRCP & $\gO(N_\mu^2(N_r+N_bN_k))$\\
		Construct Hamiltonian & Matrix multiplication & $\gO(N_\mu^2N_b N_k)$\\
		\multirow{2}{*}{Fourier interpolation} & Fast Fourier transform (FFT) & $\gO(N_\mu^2N_k\log(N_k))$ \\
		& Nonuniform FFT (NUFFT) or butterfly factorization\cite{li2015butterfly} & $\gO(N_\mu^2N_q \log(N_q))$ \\
		Diagonalization & Iterative diagonalization & $\gO(N_\mu^2N_bN_q)$\\
		\hline \hline
	\end{tabular*}
	\label{tab:complexity}
\end{table}

\begin{figure*}[htbp]
	\begin{tikzpicture}
		\node[anchor=center] at (0,0) {\includegraphics[width=\textwidth]{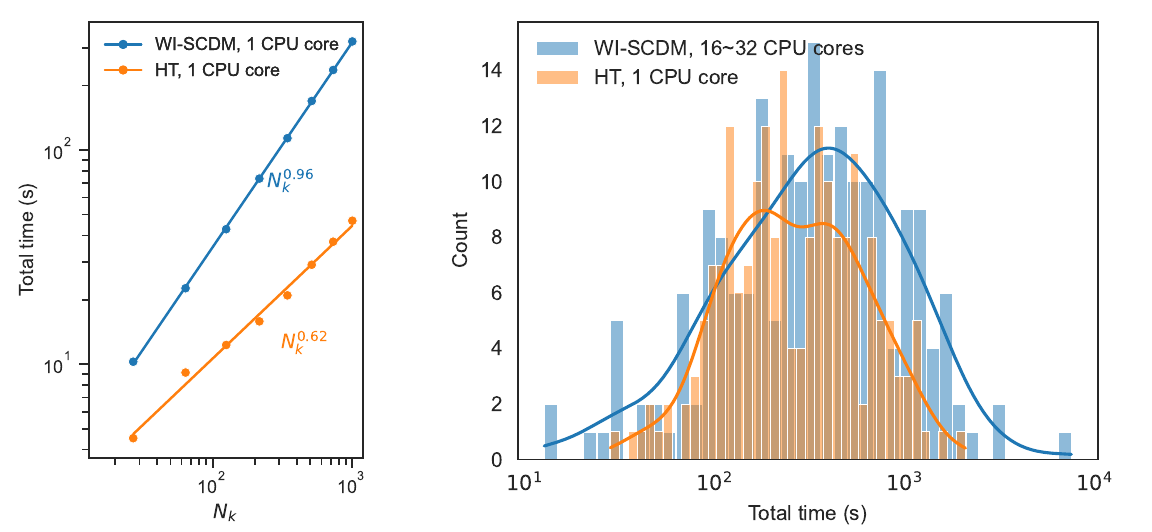}};
		\node[anchor=center] at (-6.1,2.8) {(a)};
		\node[anchor=center] at (-1.2,2.8) {(b)};
	\end{tikzpicture}
	\caption{Computational scaling and timing comparison of HT versus WI-SCDM.
		(a) Computational time as a function of $N_k$ for HT and WI-SCDM on the Si$_8$ system, performed on a single CPU core. Despite using a larger basis set, HT demonstrates faster performance and a lower scaling compared to WI-SCDM.
		(b) Actual computational time in high-throughput calculations for HT and WI-SCDM. HT runs on a single CPU core, while WI-SCDM utilizes 16 and 32 CPU cores for different tasks. HT is more efficient for large systems, whereas WI-SCDM performs better for smaller systems.}
	\label{fig:efficiency}
\end{figure*}

We perform tests on the Si$_8$ system by varying $N_k$ to compare the time complexity of HT and WI-SCDM. The tests are conducted on a single CPU core with parallelization disabled.
In Fig.~\ref{fig:efficiency}(a), although HT uses a larger basis set, it is still faster, requiring less computational time and exhibiting a lower scaling of $N_k^{0.62}$.
In contrast, WI-SCDM requires run-time optimization, making it slower and showing a scaling of $N_k^{0.96}$.
Theoretically, HT is expected to scale linearly with $N_k$, but we observe sublinear scaling.
The reason is that the key computational steps of HT depend on the size of numerical basis set $N_\mu$ instead of $N_k$, and $N_\mu$ scales sublinearly with respect to $N_k$.
Specifically, as $N_k$ approaches infinity, $N_\mu$ tends toward a constant.
Additional tests on $N_\mu$ are provided in Supplementary Material S2.
We expect that when $N_k$ becomes large enough, the steps that scale linearly with $N_k$ will dominate the computational time of HT, causing the observed results to align with the theoretical scaling.

Furthermore, we present the computational time for both HT and WI-SCDM in the high-throughput calculations, as shown in the cumulative frequency histogram of Fig.~\ref{fig:efficiency}(b).
Currently, HT does not support MPI parallelization and runs on a single CPU core.
The WI-SCDM calculations use 16 CPU cores for computing the overlap and projection matrices with pw2wannier90.x, and 32 CPU cores for constructing MLWFs with wannier90.x.
The runtime for both methods typically falls between $10^2$ and $10^3$ seconds, with WI-SCDM being faster for small systems but slower for larger ones. In HT, the primary bottleneck is the construction of overlap matrices and the explicit Hamiltonian when using the PAW method, which accounts for more than 50\% of the total time.

\section{Discussion}
The localization of the Hamiltonian is the primary factor influencing interpolation accuracy. HT eliminates the need for the complex runtime optimization procedures required in WI by directly localizing the Hamiltonian through a pre-optimized eigenvalue transformation.
By employing this transformation, HT could restore the localization of the Hamiltonian and achieve significantly higher accuracy than WI-SCDM.
In our tests, HT demonstrates superior performance in handling entangled bands and GW quasiparticle band structures, providing both improved accuracy and efficiency. HT offers a robust and efficient alternative to WI-SCDM, particularly for complex electronic structure calculations.
Moreover, WI-SCDM-f, which integrates the transform function $f$ with the WI-SCDM method, produces model Hamiltonians that are more accurate than those obtained by WI-SCDM alone.

\section{Methods}
\subsection{Code implementation}
The HT method is implemented in Quantum ESPRESSO (QE)\cite{giannozzi2009quantum, giannozzi2017advanced, giannozzi2020quantum}.
Currently, NUFFT and iterative diagonalization are not yet implemented in the code; they are temporarily replaced by matrix multiplication and direct diagonalization, respectively.
DFT calculations are performed using QE with the Perdew-Burke-Ernzerhof (PBE) functional within the generalized gradient approximation (GGA)~\cite{perdew1996generalized}.
Quasi-particle energies at the GW level are computed using BerkeleyGW\cite{hybertsen1986electron, deslippe2012berkeleygw}.
Wannier interpolations are performed with Wannier90~\cite{pizzi2020wannier90}.

\subsection{Parameters of calculation}
In the high-throughput calculations, pseudopotentials from the SSSP efficiency library (version 1.1, PBE functional)~\cite{prandini2018precision} are used, along with the recommended energy cutoffs.
The $\vk$-point mesh is chosen with a spacing of 0.2 \AA$^{-1}$.
For other DFT calculations, the optimized norm-conserving Vanderbilt (ONCV) pseudopotentials~\cite{schlipf2015optimization} are used.
In the test of Fig.~\ref{fig:accuracy}(b), we use a cutoff energy of 25 Ry, and sp$^3$ projections for constructing MLWFs.
In the test of Fig.~\ref{fig:accuracy}(c), cutoff energy is 100 Ry, SCDM-$\mu$ is 10, SCDM-$\sigma$ is 2.

\backmatter
\bmhead{Acknowledgments}
We thank Anil Damle, Yong Xu and Zhao Liu for valuable comments.
This work is partly supported by the National Natural Science Foundation of China (22173093, 21688102, 12271109), the National Key
Research and Development Program of China (2016YFA0200604, 2021YFB0300600), the Anhui Provincial Key Research and Development Program (2022a05020052), the Anhui Initiative in Quantum Information Technologies (AHY090400), the CAS Project for Young Scientists in Basic Research (YSBR-005), by the Hefei National Laboratory for Physical Sciences at the Microscale (KF2020003). Lin Lin is a Simons Investigator. We thank the Supercomputing Center of Chinese Academy of Sciences, the Supercomputing Center of USTC, the National Supercomputing Center in Wuxi, and Tianjin, Shanghai, and Guangzhou Supercomputing Centers for the computational resources.

\bmhead{Data availability}
The test data are available in the Materials Cloud Archive at \url{https://doi.org/10.24435/materialscloud:v1-52}.

\bmhead{Code availability}
The HT code is available at \url{https://github.com/KaiWu0/Hamiltonian_Transformation}.

\bibliography{reference}


\begin{thebibliography}{37}
\ifx \bisbn   \undefined \def \bisbn  #1{ISBN #1}\fi
\ifx \binits  \undefined \def \binits#1{#1}\fi
\ifx \bauthor  \undefined \def \bauthor#1{#1}\fi
\ifx \batitle  \undefined \def \batitle#1{#1}\fi
\ifx \bjtitle  \undefined \def \bjtitle#1{#1}\fi
\ifx \bvolume  \undefined \def \bvolume#1{\textbf{#1}}\fi
\ifx \byear  \undefined \def \byear#1{#1}\fi
\ifx \bissue  \undefined \def \bissue#1{#1}\fi
\ifx \bfpage  \undefined \def \bfpage#1{#1}\fi
\ifx \blpage  \undefined \def \blpage #1{#1}\fi
\ifx \burl  \undefined \def \burl#1{\textsf{#1}}\fi
\ifx \doiurl  \undefined \def \doiurl#1{\url{https://doi.org/#1}}\fi
\ifx \betal  \undefined \def \betal{\textit{et al.}}\fi
\ifx \binstitute  \undefined \def \binstitute#1{#1}\fi
\ifx \binstitutionaled  \undefined \def \binstitutionaled#1{#1}\fi
\ifx \bctitle  \undefined \def \bctitle#1{#1}\fi
\ifx \beditor  \undefined \def \beditor#1{#1}\fi
\ifx \bpublisher  \undefined \def \bpublisher#1{#1}\fi
\ifx \bbtitle  \undefined \def \bbtitle#1{#1}\fi
\ifx \bedition  \undefined \def \bedition#1{#1}\fi
\ifx \bseriesno  \undefined \def \bseriesno#1{#1}\fi
\ifx \blocation  \undefined \def \blocation#1{#1}\fi
\ifx \bsertitle  \undefined \def \bsertitle#1{#1}\fi
\ifx \bsnm \undefined \def \bsnm#1{#1}\fi
\ifx \bsuffix \undefined \def \bsuffix#1{#1}\fi
\ifx \bparticle \undefined \def \bparticle#1{#1}\fi
\ifx \barticle \undefined \def \barticle#1{#1}\fi
\bibcommenthead
\ifx \bconfdate \undefined \def \bconfdate #1{#1}\fi
\ifx \botherref \undefined \def \botherref #1{#1}\fi
\ifx \url \undefined \def \url#1{\textsf{#1}}\fi
\ifx \bchapter \undefined \def \bchapter#1{#1}\fi
\ifx \bbook \undefined \def \bbook#1{#1}\fi
\ifx \bcomment \undefined \def \bcomment#1{#1}\fi
\ifx \oauthor \undefined \def \oauthor#1{#1}\fi
\ifx \citeauthoryear \undefined \def \citeauthoryear#1{#1}\fi
\ifx \endbibitem  \undefined \def \endbibitem {}\fi
\ifx \bconflocation  \undefined \def \bconflocation#1{#1}\fi
\ifx \arxivurl  \undefined \def \arxivurl#1{\textsf{#1}}\fi
\csname PreBibitemsHook\endcsname

\bibitem[\protect\citeauthoryear{Hohenberg and
  Kohn}{1964}]{hohenberg1964inhomogeneous}
\begin{barticle}
\bauthor{\bsnm{Hohenberg}, \binits{P.}},
\bauthor{\bsnm{Kohn}, \binits{W.}}:
\batitle{{Inhomogeneous electron gas}}.
\bjtitle{Phys. Rev.}
\bvolume{136}(\bissue{3B}),
\bfpage{864}
(\byear{1964})
\end{barticle}
\endbibitem

\bibitem[\protect\citeauthoryear{Kohn and Sham}{1965}]{kohn1965self}
\begin{barticle}
\bauthor{\bsnm{Kohn}, \binits{W.}},
\bauthor{\bsnm{Sham}, \binits{L.J.}}:
\batitle{{Self-consistent equations including exchange and correlation
  effects}}.
\bjtitle{Phys. Rev.}
\bvolume{140}(\bissue{4A}),
\bfpage{1133}
(\byear{1965})
\end{barticle}
\endbibitem

\bibitem[\protect\citeauthoryear{Marzari and
  Vanderbilt}{1997}]{marzari1997maximally}
\begin{barticle}
\bauthor{\bsnm{Marzari}, \binits{N.}},
\bauthor{\bsnm{Vanderbilt}, \binits{D.}}:
\batitle{{Maximally localized generalized Wannier functions for composite
  energy bands}}.
\bjtitle{Phys. Rev. B}
\bvolume{56}(\bissue{20}),
\bfpage{12847}
(\byear{1997})
\end{barticle}
\endbibitem

\bibitem[\protect\citeauthoryear{Marzari et~al.}{2012}]{marzari2012maximally}
\begin{barticle}
\bauthor{\bsnm{Marzari}, \binits{N.}},
\bauthor{\bsnm{Mostofi}, \binits{A.A.}},
\bauthor{\bsnm{Yates}, \binits{J.R.}},
\bauthor{\bsnm{Souza}, \binits{I.}},
\bauthor{\bsnm{Vanderbilt}, \binits{D.}}:
\batitle{{Maximally localized Wannier functions: Theory and applications}}.
\bjtitle{Rev. Mod. Phys.}
\bvolume{84}(\bissue{4}),
\bfpage{1419}
(\byear{2012})
\end{barticle}
\endbibitem

\bibitem[\protect\citeauthoryear{Pizzi et~al.}{2020}]{pizzi2020wannier90}
\begin{barticle}
\bauthor{\bsnm{Pizzi}, \binits{G.}},
\bauthor{\bsnm{Vitale}, \binits{V.}},
\bauthor{\bsnm{Arita}, \binits{R.}},
\bauthor{\bsnm{Bl{\"u}gel}, \binits{S.}},
\bauthor{\bsnm{Freimuth}, \binits{F.}},
\bauthor{\bsnm{G{\'e}ranton}, \binits{G.}},
\bauthor{\bsnm{Gibertini}, \binits{M.}},
\bauthor{\bsnm{Gresch}, \binits{D.}},
\bauthor{\bsnm{Johnson}, \binits{C.}},
\bauthor{\bsnm{Koretsune}, \binits{T.}}, \betal:
\batitle{{Wannier90 as a community code: new features and applications}}.
\bjtitle{J. Phys. Condens. Matter}
\bvolume{32}(\bissue{16}),
\bfpage{165902}
(\byear{2020})
\end{barticle}
\endbibitem

\bibitem[\protect\citeauthoryear{Jung and MacDonald}{2013}]{jung2013tight}
\begin{barticle}
\bauthor{\bsnm{Jung}, \binits{J.}},
\bauthor{\bsnm{MacDonald}, \binits{A.H.}}:
\batitle{{Tight-binding model for graphene $\pi$-bands from maximally localized
  Wannier functions}}.
\bjtitle{Phys. Rev. B}
\bvolume{87}(\bissue{19}),
\bfpage{195450}
(\byear{2013})
\end{barticle}
\endbibitem

\bibitem[\protect\citeauthoryear{Garrity and
  Choudhary}{2021}]{garrity2021database}
\begin{barticle}
\bauthor{\bsnm{Garrity}, \binits{K.F.}},
\bauthor{\bsnm{Choudhary}, \binits{K.}}:
\batitle{{Database of Wannier tight-binding Hamiltonians using high-throughput
  density functional theory}}.
\bjtitle{Sci. Data}
\bvolume{8}(\bissue{1}),
\bfpage{1}--\blpage{10}
(\byear{2021})
\end{barticle}
\endbibitem

\bibitem[\protect\citeauthoryear{Wang et~al.}{2006}]{wang2006ab}
\begin{barticle}
\bauthor{\bsnm{Wang}, \binits{X.}},
\bauthor{\bsnm{Yates}, \binits{J.R.}},
\bauthor{\bsnm{Souza}, \binits{I.}},
\bauthor{\bsnm{Vanderbilt}, \binits{D.}}:
\batitle{{Ab initio calculation of the anomalous Hall conductivity by Wannier
  interpolation}}.
\bjtitle{Phys. Rev. B}
\bvolume{74}(\bissue{19}),
\bfpage{195118}
(\byear{2006})
\end{barticle}
\endbibitem

\bibitem[\protect\citeauthoryear{Yates et~al.}{2007}]{yates2007spectral}
\begin{barticle}
\bauthor{\bsnm{Yates}, \binits{J.R.}},
\bauthor{\bsnm{Wang}, \binits{X.}},
\bauthor{\bsnm{Vanderbilt}, \binits{D.}},
\bauthor{\bsnm{Souza}, \binits{I.}}:
\batitle{{Spectral and Fermi surface properties from Wannier interpolation}}.
\bjtitle{Phys. Rev. B}
\bvolume{75}(\bissue{19}),
\bfpage{195121}
(\byear{2007})
\end{barticle}
\endbibitem

\bibitem[\protect\citeauthoryear{Wang et~al.}{2017}]{wang2017first}
\begin{barticle}
\bauthor{\bsnm{Wang}, \binits{C.}},
\bauthor{\bsnm{Liu}, \binits{X.}},
\bauthor{\bsnm{Kang}, \binits{L.}},
\bauthor{\bsnm{Gu}, \binits{B.-L.}},
\bauthor{\bsnm{Xu}, \binits{Y.}},
\bauthor{\bsnm{Duan}, \binits{W.}}:
\batitle{{First-principles calculation of nonlinear optical responses by
  Wannier interpolation}}.
\bjtitle{Phys. Rev. B}
\bvolume{96}(\bissue{11}),
\bfpage{115147}
(\byear{2017})
\end{barticle}
\endbibitem

\bibitem[\protect\citeauthoryear{Mustafa
  et~al.}{2015}]{MustafaCohCohenEtAl2015}
\begin{barticle}
\bauthor{\bsnm{Mustafa}, \binits{J.I.}},
\bauthor{\bsnm{Coh}, \binits{S.}},
\bauthor{\bsnm{Cohen}, \binits{M.L.}},
\bauthor{\bsnm{Louie}, \binits{S.G.}}:
\batitle{{Automated construction of maximally localized Wannier functions:
  Optimized projection functions method}}.
\bjtitle{Phys. Rev. B}
\bvolume{92},
\bfpage{165134}
(\byear{2015})
\end{barticle}
\endbibitem

\bibitem[\protect\citeauthoryear{Canc{\`e}s
  et~al.}{2017}]{CancesLevittPanatiEtAl2017}
\begin{barticle}
\bauthor{\bsnm{Canc{\`e}s}, \binits{E.}},
\bauthor{\bsnm{Levitt}, \binits{A.}},
\bauthor{\bsnm{Panati}, \binits{G.}},
\bauthor{\bsnm{Stoltz}, \binits{G.}}:
\batitle{Robust determination of maximally-localized {Wannier} functions}.
\bjtitle{Phys. Rev. B}
\bvolume{95},
\bfpage{075114}
(\byear{2017})
\end{barticle}
\endbibitem

\bibitem[\protect\citeauthoryear{Stubbs et~al.}{2021}]{StubbsWatsonLu2021}
\begin{barticle}
\bauthor{\bsnm{Stubbs}, \binits{K.D.}},
\bauthor{\bsnm{Watson}, \binits{A.B.}},
\bauthor{\bsnm{Lu}, \binits{J.}}:
\batitle{{Iterated projected position algorithm for constructing exponentially
  localized generalized Wannier functions for periodic and nonperiodic
  insulators in two dimensions and higher}}.
\bjtitle{Phys. Rev. B}
\bvolume{103}(\bissue{7}),
\bfpage{075125}
(\byear{2021})
\end{barticle}
\endbibitem

\bibitem[\protect\citeauthoryear{Qiao et~al.}{2023}]{qiao2023projectability}
\begin{barticle}
\bauthor{\bsnm{Qiao}, \binits{J.}},
\bauthor{\bsnm{Pizzi}, \binits{G.}},
\bauthor{\bsnm{Marzari}, \binits{N.}}:
\batitle{Projectability disentanglement for accurate and automated
  electronic-structure hamiltonians}.
\bjtitle{Npj Comput. Mater.}
\bvolume{9}(\bissue{1}),
\bfpage{208}
(\byear{2023})
\end{barticle}
\endbibitem

\bibitem[\protect\citeauthoryear{Damle et~al.}{2015}]{damle2015compressed}
\begin{barticle}
\bauthor{\bsnm{Damle}, \binits{A.}},
\bauthor{\bsnm{Lin}, \binits{L.}},
\bauthor{\bsnm{Ying}, \binits{L.}}:
\batitle{{Compressed representation of Kohn--Sham orbitals via selected columns
  of the density matrix}}.
\bjtitle{J. Chem. Theory Comput.}
\bvolume{11}(\bissue{4}),
\bfpage{1463}--\blpage{1469}
(\byear{2015})
\end{barticle}
\endbibitem

\bibitem[\protect\citeauthoryear{Damle et~al.}{2017}]{damle2017scdm}
\begin{barticle}
\bauthor{\bsnm{Damle}, \binits{A.}},
\bauthor{\bsnm{Lin}, \binits{L.}},
\bauthor{\bsnm{Ying}, \binits{L.}}:
\batitle{{SCDM-k: Localized orbitals for solids via selected columns of the
  density matrix}}.
\bjtitle{J. Comput. Phys.}
\bvolume{334},
\bfpage{1}--\blpage{15}
(\byear{2017})
\end{barticle}
\endbibitem

\bibitem[\protect\citeauthoryear{Damle and
  Lin}{2018}]{damle2018disentanglement}
\begin{barticle}
\bauthor{\bsnm{Damle}, \binits{A.}},
\bauthor{\bsnm{Lin}, \binits{L.}}:
\batitle{{Disentanglement via entanglement: a unified method for Wannier
  localization}}.
\bjtitle{Multiscale Model. Simul.}
\bvolume{16}(\bissue{3}),
\bfpage{1392}--\blpage{1410}
(\byear{2018})
\end{barticle}
\endbibitem

\bibitem[\protect\citeauthoryear{Soluyanov and
  Vanderbilt}{2011}]{soluyanov2011wannier}
\begin{barticle}
\bauthor{\bsnm{Soluyanov}, \binits{A.A.}},
\bauthor{\bsnm{Vanderbilt}, \binits{D.}}:
\batitle{{Wannier representation of Z$_2$ topological insulators}}.
\bjtitle{Phys. Rev. B}
\bvolume{83}(\bissue{3}),
\bfpage{035108}
(\byear{2011})
\end{barticle}
\endbibitem

\bibitem[\protect\citeauthoryear{Cornean et~al.}{2017}]{cornean2017wannier}
\begin{barticle}
\bauthor{\bsnm{Cornean}, \binits{H.D.}},
\bauthor{\bsnm{Monaco}, \binits{D.}},
\bauthor{\bsnm{Teufel}, \binits{S.}}:
\batitle{{Wannier functions and Z$_2$ invariants in time-reversal symmetric
  topological insulators}}.
\bjtitle{Rev. Math. Phys.}
\bvolume{29}(\bissue{02}),
\bfpage{1730001}
(\byear{2017})
\end{barticle}
\endbibitem

\bibitem[\protect\citeauthoryear{Souza
  et~al.}{2001}]{SouzaMarzariVanderbilt2001}
\begin{barticle}
\bauthor{\bsnm{Souza}, \binits{I.}},
\bauthor{\bsnm{Marzari}, \binits{N.}},
\bauthor{\bsnm{Vanderbilt}, \binits{D.}}:
\batitle{Maximally localized wannier functions for entangled energy bands}.
\bjtitle{Phys. Rev. B}
\bvolume{65},
\bfpage{035109}
(\byear{2001})
\end{barticle}
\endbibitem

\bibitem[\protect\citeauthoryear{Damle et~al.}{2019}]{damle2019variational}
\begin{barticle}
\bauthor{\bsnm{Damle}, \binits{A.}},
\bauthor{\bsnm{Levitt}, \binits{A.}},
\bauthor{\bsnm{Lin}, \binits{L.}}:
\batitle{{Variational formulation for Wannier functions with entangled band
  structure}}.
\bjtitle{Multiscale Model. Simul.}
\bvolume{17}(\bissue{1}),
\bfpage{167}--\blpage{191}
(\byear{2019})
\end{barticle}
\endbibitem

\bibitem[\protect\citeauthoryear{Pasquini and
  Reichel}{2006}]{pasquini2006tridiagonal}
\begin{barticle}
\bauthor{\bsnm{Pasquini}, \binits{S.N.L.}},
\bauthor{\bsnm{Reichel}, \binits{L.}}:
\batitle{{Tridiagonal {Toeplitz} matrices: Properties and novel applications}}.
\bjtitle{Numer. Linear Algebra Appl.}
\bvolume{30},
\bfpage{302}--\blpage{326}
(\byear{2006})
\end{barticle}
\endbibitem

\bibitem[\protect\citeauthoryear{Baer and Head-Gordon}{1997}]{baer1997sparsity}
\begin{barticle}
\bauthor{\bsnm{Baer}, \binits{R.}},
\bauthor{\bsnm{Head-Gordon}, \binits{M.}}:
\batitle{{Sparsity of the density matrix in Kohn-Sham density functional theory
  and an assessment of linear system-size scaling methods}}.
\bjtitle{Phys. Rev. Lett.}
\bvolume{79}(\bissue{20}),
\bfpage{3962}
(\byear{1997})
\end{barticle}
\endbibitem

\bibitem[\protect\citeauthoryear{Benzi et~al.}{2013}]{benzi2013decay}
\begin{barticle}
\bauthor{\bsnm{Benzi}, \binits{M.}},
\bauthor{\bsnm{Boito}, \binits{P.}},
\bauthor{\bsnm{Razouk}, \binits{N.}}:
\batitle{{Decay properties of spectral projectors with applications to
  electronic structure}}.
\bjtitle{SIAM Rev.}
\bvolume{55}(\bissue{1}),
\bfpage{3}--\blpage{64}
(\byear{2013})
\end{barticle}
\endbibitem

\bibitem[\protect\citeauthoryear{Bernstein}{1912}]{bernstein1912ordre}
\begin{bbook}
\bauthor{\bsnm{Bernstein}, \binits{S.}}:
\bbtitle{{Sur L'ordre de la Meilleure Approximation des Fonctions Continues Par
  des Polyn{\^o}mes de Degr{\'e} Donn{\'e}}}
vol. \bseriesno{7},
(\byear{1912})
\end{bbook}
\endbibitem

\bibitem[\protect\citeauthoryear{Xiang et~al.}{2010}]{xiang2010error}
\begin{barticle}
\bauthor{\bsnm{Xiang}, \binits{S.}},
\bauthor{\bsnm{Chen}, \binits{X.}},
\bauthor{\bsnm{Wang}, \binits{H.}}:
\batitle{{Error bounds for approximation in Chebyshev points}}.
\bjtitle{Numer Math (Heidelb)}
\bvolume{116}(\bissue{3}),
\bfpage{463}--\blpage{491}
(\byear{2010})
\end{barticle}
\endbibitem

\bibitem[\protect\citeauthoryear{Duersch and Gu}{2017}]{duersch2017randomized}
\begin{barticle}
\bauthor{\bsnm{Duersch}, \binits{J.A.}},
\bauthor{\bsnm{Gu}, \binits{M.}}:
\batitle{Randomized {QR} with column pivoting}.
\bjtitle{SIAM J. Sci. Comput.}
\bvolume{39}(\bissue{4}),
\bfpage{263}--\blpage{291}
(\byear{2017})
\end{barticle}
\endbibitem

\bibitem[\protect\citeauthoryear{Vitale et~al.}{2020}]{vitale2020automated}
\begin{barticle}
\bauthor{\bsnm{Vitale}, \binits{V.}},
\bauthor{\bsnm{Pizzi}, \binits{G.}},
\bauthor{\bsnm{Marrazzo}, \binits{A.}},
\bauthor{\bsnm{Yates}, \binits{J.R.}},
\bauthor{\bsnm{Marzari}, \binits{N.}},
\bauthor{\bsnm{Mostofi}, \binits{A.A.}}:
\batitle{Automated high-throughput wannierisation}.
\bjtitle{Npj Comput. Mater.}
\bvolume{6}(\bissue{1}),
\bfpage{66}
(\byear{2020})
\end{barticle}
\endbibitem

\bibitem[\protect\citeauthoryear{Deslippe
  et~al.}{2012}]{deslippe2012berkeleygw}
\begin{barticle}
\bauthor{\bsnm{Deslippe}, \binits{J.}},
\bauthor{\bsnm{Samsonidze}, \binits{G.}},
\bauthor{\bsnm{Strubbe}, \binits{D.A.}},
\bauthor{\bsnm{Jain}, \binits{M.}},
\bauthor{\bsnm{Cohen}, \binits{M.L.}},
\bauthor{\bsnm{Louie}, \binits{S.G.}}:
\batitle{{BerkeleyGW: A massively parallel computer package for the calculation
  of the quasiparticle and optical properties of materials and
  nanostructures}}.
\bjtitle{Comput. Phys. Commun.}
\bvolume{183}(\bissue{6}),
\bfpage{1269}--\blpage{1289}
(\byear{2012})
\end{barticle}
\endbibitem

\bibitem[\protect\citeauthoryear{Li et~al.}{2015}]{li2015butterfly}
\begin{barticle}
\bauthor{\bsnm{Li}, \binits{Y.}},
\bauthor{\bsnm{Yang}, \binits{H.}},
\bauthor{\bsnm{Martin}, \binits{E.R.}},
\bauthor{\bsnm{Ho}, \binits{K.L.}},
\bauthor{\bsnm{Ying}, \binits{L.}}:
\batitle{Butterfly factorization}.
\bjtitle{Multiscale Model. Simul.}
\bvolume{13}(\bissue{2}),
\bfpage{714}--\blpage{732}
(\byear{2015})
\end{barticle}
\endbibitem

\bibitem[\protect\citeauthoryear{Giannozzi et~al.}{2009}]{giannozzi2009quantum}
\begin{barticle}
\bauthor{\bsnm{Giannozzi}, \binits{P.}},
\bauthor{\bsnm{Baroni}, \binits{S.}},
\bauthor{\bsnm{Bonini}, \binits{N.}},
\bauthor{\bsnm{Calandra}, \binits{M.}},
\bauthor{\bsnm{Car}, \binits{R.}},
\bauthor{\bsnm{Cavazzoni}, \binits{C.}},
\bauthor{\bsnm{Ceresoli}, \binits{D.}},
\bauthor{\bsnm{Chiarotti}, \binits{G.L.}},
\bauthor{\bsnm{Cococcioni}, \binits{M.}},
\bauthor{\bsnm{Dabo}, \binits{I.}}, \betal:
\batitle{{QUANTUM ESPRESSO: a modular and open-source software project for
  quantum simulations of materials}}.
\bjtitle{J. Phys.: Condens. Matter}
\bvolume{21}(\bissue{39}),
\bfpage{395502}
(\byear{2009})
\end{barticle}
\endbibitem

\bibitem[\protect\citeauthoryear{Giannozzi
  et~al.}{2017}]{giannozzi2017advanced}
\begin{barticle}
\bauthor{\bsnm{Giannozzi}, \binits{P.}},
\bauthor{\bsnm{Andreussi}, \binits{O.}},
\bauthor{\bsnm{Brumme}, \binits{T.}},
\bauthor{\bsnm{Bunau}, \binits{O.}},
\bauthor{\bsnm{Nardelli}, \binits{M.B.}},
\bauthor{\bsnm{Calandra}, \binits{M.}},
\bauthor{\bsnm{Car}, \binits{R.}},
\bauthor{\bsnm{Cavazzoni}, \binits{C.}},
\bauthor{\bsnm{Ceresoli}, \binits{D.}},
\bauthor{\bsnm{Cococcioni}, \binits{M.}}, \betal:
\batitle{{Advanced capabilities for materials modelling with Quantum
  ESPRESSO}}.
\bjtitle{J. Phys.: Condens. Matter}
\bvolume{29}(\bissue{46}),
\bfpage{465901}
(\byear{2017})
\end{barticle}
\endbibitem

\bibitem[\protect\citeauthoryear{Giannozzi et~al.}{2020}]{giannozzi2020quantum}
\begin{barticle}
\bauthor{\bsnm{Giannozzi}, \binits{P.}},
\bauthor{\bsnm{Baseggio}, \binits{O.}},
\bauthor{\bsnm{Bonf{\`a}}, \binits{P.}},
\bauthor{\bsnm{Brunato}, \binits{D.}},
\bauthor{\bsnm{Car}, \binits{R.}},
\bauthor{\bsnm{Carnimeo}, \binits{I.}},
\bauthor{\bsnm{Cavazzoni}, \binits{C.}},
\bauthor{\bsnm{De~Gironcoli}, \binits{S.}},
\bauthor{\bsnm{Delugas}, \binits{P.}},
\bauthor{\bsnm{Ferrari~Ruffino}, \binits{F.}}, \betal:
\batitle{{Quantum ESPRESSO toward the exascale}}.
\bjtitle{J. Chem. Phys.}
\bvolume{152}(\bissue{15}),
\bfpage{154105}
(\byear{2020})
\end{barticle}
\endbibitem

\bibitem[\protect\citeauthoryear{Perdew et~al.}{1996}]{perdew1996generalized}
\begin{barticle}
\bauthor{\bsnm{Perdew}, \binits{J.P.}},
\bauthor{\bsnm{Burke}, \binits{K.}},
\bauthor{\bsnm{Ernzerhof}, \binits{M.}}:
\batitle{{Generalized gradient approximation made simple}}.
\bjtitle{Phys. Rev. Lett.}
\bvolume{77}(\bissue{18}),
\bfpage{3865}
(\byear{1996})
\end{barticle}
\endbibitem

\bibitem[\protect\citeauthoryear{Hybertsen and
  Louie}{1986}]{hybertsen1986electron}
\begin{barticle}
\bauthor{\bsnm{Hybertsen}, \binits{M.S.}},
\bauthor{\bsnm{Louie}, \binits{S.G.}}:
\batitle{{Electron correlation in semiconductors and insulators: Band gaps and
  quasiparticle energies}}.
\bjtitle{Phys. Rev. B}
\bvolume{34}(\bissue{8}),
\bfpage{5390}
(\byear{1986})
\end{barticle}
\endbibitem

\bibitem[\protect\citeauthoryear{Prandini et~al.}{2018}]{prandini2018precision}
\begin{barticle}
\bauthor{\bsnm{Prandini}, \binits{G.}},
\bauthor{\bsnm{Marrazzo}, \binits{A.}},
\bauthor{\bsnm{Castelli}, \binits{I.E.}},
\bauthor{\bsnm{Mounet}, \binits{N.}},
\bauthor{\bsnm{Marzari}, \binits{N.}}:
\batitle{Precision and efficiency in solid-state pseudopotential calculations}.
\bjtitle{Npj Comput. Mater.}
\bvolume{4}(\bissue{1}),
\bfpage{72}
(\byear{2018})
\end{barticle}
\endbibitem

\bibitem[\protect\citeauthoryear{Schlipf and
  Gygi}{2015}]{schlipf2015optimization}
\begin{barticle}
\bauthor{\bsnm{Schlipf}, \binits{M.}},
\bauthor{\bsnm{Gygi}, \binits{F.}}:
\batitle{{Optimization algorithm for the generation of ONCV pseudopotentials}}.
\bjtitle{Comput. Phys. Commun.}
\bvolume{196},
\bfpage{36}--\blpage{44}
(\byear{2015})
\end{barticle}
\endbibitem

\end{thebibliography}
\end{document}